\documentclass[]{elsart}
\usepackage{graphicx}

\usepackage{amssymb}

\usepackage{natbib}

\begin{document}

\begin{frontmatter}



\title{Synthesis and Properties of YbB$_2$}


\author{M. A. Avila\corauthref{cor1}}
\corauth[cor1]{Corresponding author}
\ead{avila@ameslab.gov},
\author{S. L. Bud'ko},
\author{C. Petrovic\thanksref{presad}}
\thanks[presad]{\textit{Present address:} Brookhaven National Laboratory, Upton, NY 11973},
\author{R. A. Ribeiro},
\author{P. C. Canfield}%
\address{Ames Laboratory and Department of Physics and Astronomy\\
Iowa State University, Ames, IA 50011}
\author{A. V. Tsvyashchenko},
\author{L. N. Fomicheva}%
\address{Institute for High Pressure Physics\\
RAS, Troitsk, Moscow region, 142092, Russia}

\begin{abstract}
We report temperature and field dependent measurements of the
magnetic susceptibility, specific heat and resistivity of sintered
YbB$_2$ pellets, prepared via two distinct reaction routes,
utilizing different temperatures, pressures and sintering times.
Sample behavior is affected by the preparation procedure, as a
consequence of different secondary phases, most of which were
identified via x-ray diffraction. These experiments show that
YbB$_2$ is a metal with the Yb atoms in or very close to their 3+
state. YbB$_2$ appears to order anti-ferromagnetically at
$T_N=5.6\pm0.2$~K, which can be considered a relatively high
ordering temperature for an ytterbium-based intermetallic
compound.
\end{abstract}

\end{frontmatter}

\section{Introduction}
\label{sec:intro}

With the recent discovery of superconductivity below $\sim40$~K in
MgB$_2$\cite{nag01a}, interest has been renewed in the electronic
ground state and other physical properties of metal-diborides of
the simple hexagonal AlB$_2$ type structure (space group
$P6/mmm$), a family which has a large number of members, related
to the fact that the metal ions easily deform their electronic
clouds in order to fit in between two hexagonal boron
rings\cite{spe76a}. A common route used to help understand details
of the ground state of any given compound is to introduce
elemental substitutions in the lattice, and follow the trends of
its physical properties as the proportion of the substituting
element is increased. Although some successful substitutions have
been reported in the particular case of MgB$_2$, this compound has
so far shown considerable resistance towards admitting foreign
elements in its lattice. Thus, a better understanding of physical
properties and preparation procedures of isostructural compounds
may prove useful.

YbB$_2$, in particular, has been known as a compound for about 3
decades\cite{bau74a}, but presently only crystallographic and
chemical characterizations are reported\cite{spe76a,bau74a}. Given
the rich variety of interesting electronic, thermodynamic and
magnetic properties often displayed by Yb compounds such as mixed
or intermediate valency\cite{kla77a,sva00a} and heavy fermion
behavior\cite{fis92a,yat96a,tro00a}, plus the considerations
described in the previous paragraph, it seemed worthwhile to
develop a few growth techniques for YbB$_2$ and investigate this
compound in greater detail.

\section{Experimental Details}
\label{sec:details}

We obtained sintered pellets of YbB$_2$ by two different synthetic
routes. The first route, which we will call ``low-pressure
synthesis'' (LPS), was developed in Ames Laboratory and involved
sealing mixtures of both elements in a 6 cm tantalum tube under
partial argon atmosphere. The total starting mass was between 1.5
and 2 g, and an excess of Yb (20\%) was introduced to compensate
for losses associated with the high vapor pressure of Yb. Starting
reagents were Ames Laboratory Yb ingots (99.95\%) and Eagle-Picher
$^{11}$B powder (99.95\%). For sintering temperatures up to
1200$^{\circ}$C the Ta tube was subsequently sealed inside an
evacuated quartz ampoule, and this ampoule was heated in a box
furnace for the desired temperature and time, after which it was
quenched to room temperature by placing the ampoule under running
tap water. For sintering temperatures above 1200$^{\circ}$C the
sealed Ta tube was heated inside a vertical tube furnace with
flowing argon atmosphere, and after the desired sintering time was
complete the furnace was turned off and allowed to cool to room
temperature over several hours. To describe the samples prepared
through these LPS procedures we will adopt a notation of the type
TTT/HH, where TTT is the sintering temperature in $^{\circ}$C and
HH is the sintering duration in hours.

The second route was developed in the Institute for High Pressure
Physics and used high-temperature, high-pressure synthesis (HPS).
Samples were synthesized using the high pressure cell designed by
Khvostantsev \textit{et al.}\cite{khv77a}. Stoichiometric amounts
of elemental ytterbium and boron were placed in the high pressure
cell in a NaCl envelope and heated at 6 GPa pressure to
1120$^{\circ}$C for 15 minutes (sample 6 GPa) or heated at 8 GPa
pressure until melting, then cooled down and crystallized (sample
8 GPa).

Powder X-ray diffraction measurements were made with Scintag and
Philips diffractometers using a Cu $K_{\alpha}$ radiation. A Si
powder standard was added to the sample for all runs. The Si lines
have been removed from the X-ray diffraction data, leading to
apparent gaps in the powder X-ray spectra. Peak positions were
compared to those listed in the ICDD-JCPDS database.

Magnetization measurements were made on Quantum Design MPMS
systems using the standard operating modes, allowing measurements
down to $1.8$~K in temperature and up to 55 kOe or 70 kOe in
field, depending on the machine. Samples were cooled to 1.8~K
under zero applied field, after which the desired measuring field
was applied and data taken using 6 cm scan length as the
temperature was increased in steps of 0.1-0.25~K at low
temperatures, then 1-5~K at higher temperatures. Electrical
resistance measurements were performed using these same MPMS
systems operated in external device control (EDC) mode, in
conjunction with Linear Research LR400/LR700 four-probe ac
resistance bridges. Some resistance measurements were also taken
on a Quantum Design PPMS system using its own ac resistance bridge
option. The electrical contacts were placed on the samples using
4-probe geometry. Pt wires were attached to a sample surface with
Epotek H20e silver epoxy, cured at $120^{\circ}$C for $\sim30$
minutes. The irregular shape and porosity of the pellets does not
allow a reliable estimate of the material's resistivity, so only
normalized resistance is presented in this work. Specific Heat
measurements were made on the Quantum Design PPMS system with the
high-vacuum and specific heat options, allowing measurements down
to 1.9~K. The contribution of the grease addenda used to provide
thermal contact with the sample was measured separately before
each run and discounted from the measurements afterwards.

\section{Characterization of samples prepared using low-pressure synthesis}

\subsection{X-Ray Powder Diffraction}

\begin{figure}[htb]
\includegraphics[angle=0,width=32pc]{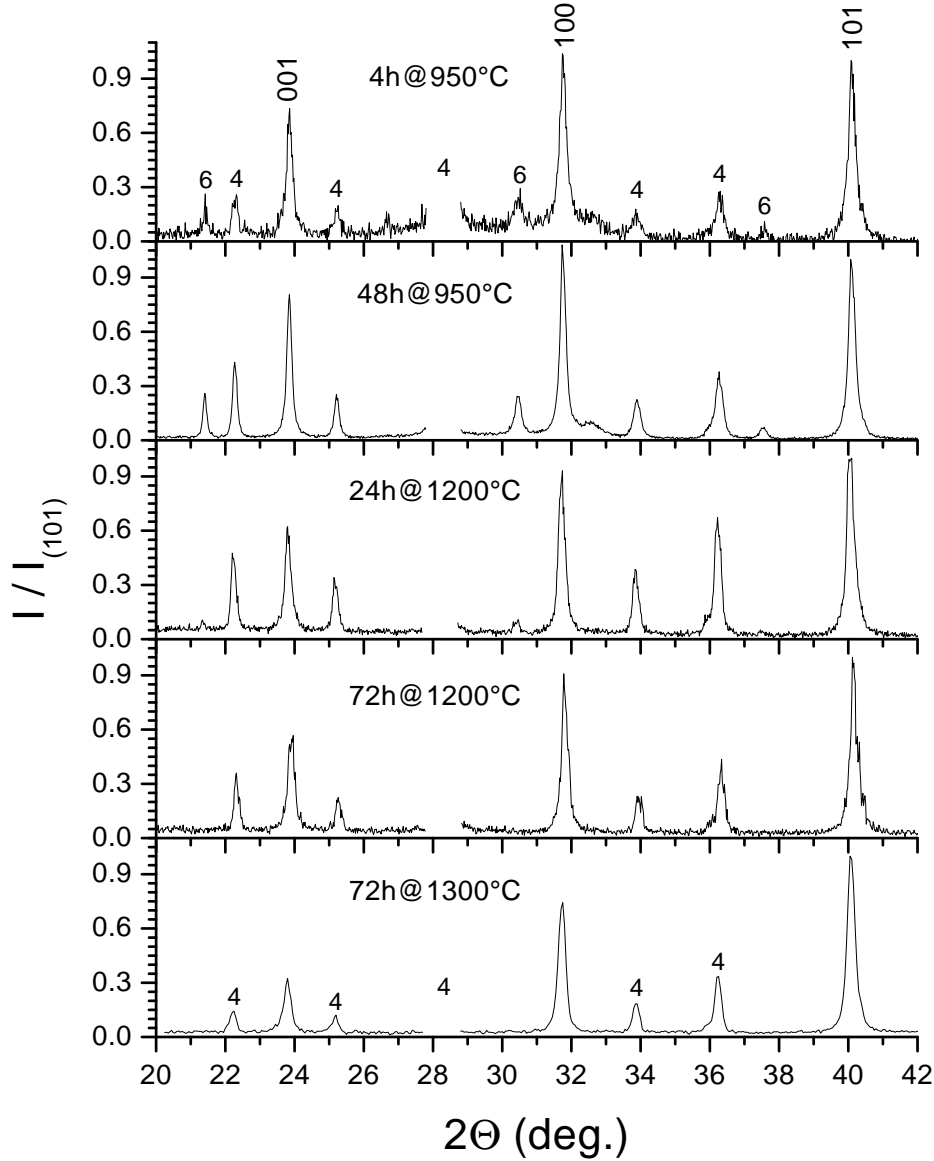}
\caption{\label{fig1} X-Ray powder diffraction pattern of YbB$_2$
from different sinterings. The 3 major peaks are the YbB$_2$
reflections in the shown interval (001,100,101), and the data has
been normalized by the intensity of the 101 peak. The other labels
in the top and bottom graphs mark the peak positions for
(4)YbB$_4$ and (6)YbB$_6$. No other compounds are detectable in
these samples.}
\end{figure}

The evolution of the YbB$_2$ sample quality with reaction
temperature and time for the LPS samples was followed with X-Ray
diffraction measurements, some of which are presented in
figure~\ref{fig1} for
$20^{\circ}\leqslant2\Theta\leqslant42^{\circ}$. The actual runs
were made between $20^{\circ}$ and $90^{\circ}$, but all the
necessary information for our analysis is contained in the region
shown.

We began by applying a procedure similar to the one we consider
ideal for sintering MgB$_2$ pellets - 950$^{\circ}$C for 4 hours,
which results in nearly 100\% pure MgB$_2$ phase. Sample 950/4
already showed mostly YbB$_2$ phase, as can be seen in the top
diffractogram of figure~\ref{fig1}, where the 3 major peaks belong
to the (001), (100) and (101) reflections of YbB$_2$. But clearly
visible are sets of peaks attributed to the higher borides YbB$_4$
and YbB$_6$, labelled as (4) and (6) respectively. A strong Si
reflection centered at 28.6$^{\circ}$ has been removed from these
data. We found no evidence for the presence of Yb$_2$O$_3$ in
these diffractograms, a common magnetic impurity in Yb-based
compounds which would manifest as a single peak above noise level
centered at 29.6$^{\circ}$. There was no clear sign of unreacted
Yb reflections either, indicating that the excess Yb introduced in
the growth mostly condenses on the internal walls of the Ta tube
which, several days after opening, turns into an oxidized powdery
layer that peels off easily. It is likely that a small amount of
Yb also condenses in the sintered pellet, but not enough to be
detectable in the diffractogram as either Yb or Yb$_2$O$_3$.

The next diffractogram in fig.~\ref{fig1} is for sample 950/48,
which was measured using more extended data acquisition time to
reduce noise level and improve resolution. No significant change
is seen in the relative heights of the measured peaks for
different phases, which would indicate that sintering for longer
times at this temperature does not significantly reduce the amount
of secondary phases in YbB$_2$.

The third, fourth and fifth diffractograms in fig.~\ref{fig1} show
that by increasing temperature and time we are able to
progressively reduce the amount of YbB$_6$ in the samples. By
1200/72 this compound is fully suppressed. However, even at
1300/72 there is still easily detectable YbB$_4$. Comparing peak
heights between different phases is not a reliable method to
quantitatively evaluate their proportion, but at a gross level we
can state from the X-Ray data that the amount of YbB$_4$ is more
than 5\% and probably less than 30\% in the 1300/72 sample. A
better estimate is obtained with the analysis of the magnetization
data below. The lattice parameters obtained from Rietveld
refinement of sample 1300/72 for the YbB$_2$ phase are
$a=3.256(1)$\AA, $c=3.735(1)$\AA, and for the YbB$_4$ phase
$a=7.071(1)$\AA, $c=3.982(1)$\AA.

\subsection{DC Magnetization}

\begin{figure}[htb]
\includegraphics[angle=0,width=32pc]{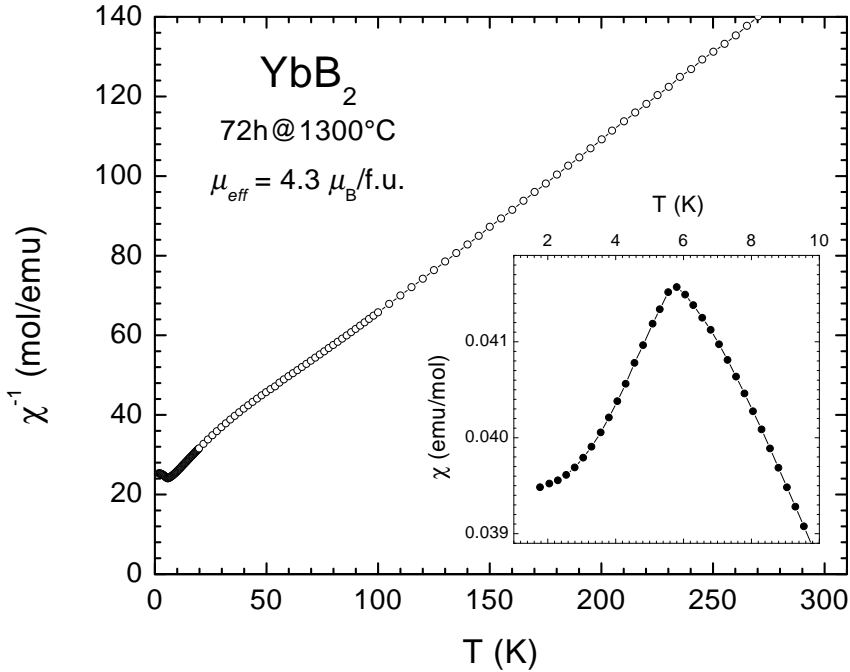}
\caption{\label{fig2} Temperature dependence of the inverse
magnetic susceptibility $\chi^{-1}(T)$ of a pellet from batch
1300/72, at an applied field of 1 kOe. The inset shows the detail
of the anti-ferromagnetic transition below 5.7~K in $\chi(T)$.}
\end{figure}

The magnetic behavior of the sintered pellets was also measured
for the different sintering procedures, and evolved consistently
with the trends seen in the X-Ray measurements. Since for both
YbB$_4$ and YbB$_6$ the Yb ions are in (or very close to) their
non-magnetic 2+ state\cite{bon78a,eto79a,tar80a}, the effect of
these impurities in magnetic measurements is essentially limited
to an overestimate of the actual value the YbB$_2$ mass, which in
turn results in underestimating the effective moment of Yb$^{3+}$
in the diboride. Indeed, by fitting the high temperature inverse
susceptibility we found that $\mu_{eff}$ increases from 3.6
$\mu_B$/YbB$_2$ for sample 950/4 to 4.3~$\mu_B$/YbB$_2$ for sample
1300/72, approaching the expected value of 4.5~$\mu_B$/Yb. By
inverting this analysis, we can estimate that about 9\% of the
1300/72 sample is YbB$_4$, assuming that it is the primary
impurity in this sample as evidenced from the X-Ray pattern. From
this point on we will focus on the results from sample 1300/72.

Figure~\ref{fig2} shows the inverse magnetic susceptibility
$\chi^{-1}(T)$ of sample 1300/72 taken in an applied field of H =
1 kOe, with the inset showing a drop in the susceptibility
$\chi(T)$ below 5.7~K, which we associate with anti-ferromagnetic
ordering of YbB$_2$. Using the criterion of maximum in $d(\chi
T)/dT$\cite{fis62a} we obtain a somewhat lower value of
$T_N=5.4\pm0.2$~K. In terms of Yb-based intermetallic compounds
this is a relatively high ordering temperature. For comparison,
Bonville \textit{et al}. had already called attention to the
unusually high ordering temperature of 4.2~K in
YbNiBC\cite{bon99a}. The high temperature inverse susceptibility
is consistent with Curie-Weiss behavior, with $\theta_p=49$~K. A
clear deviation from linear Curie-Weiss behavior is seen below
$\sim100$~K, typical of magnetic Yb compounds, and often
attributed to crystalline electric field (CEF) effects. The field
dependence of this sample's magnetization at T = 1.8~K was also
measured, it is essentially linear and featureless in the field
range of 0 - 5.5 kOe, reaching a value of 0.4~$\mu_B$/YbB$_2$ at
the highest field.

\subsection{Heat Capacity}

\begin{figure}[htb]
\includegraphics[angle=0,width=32pc]{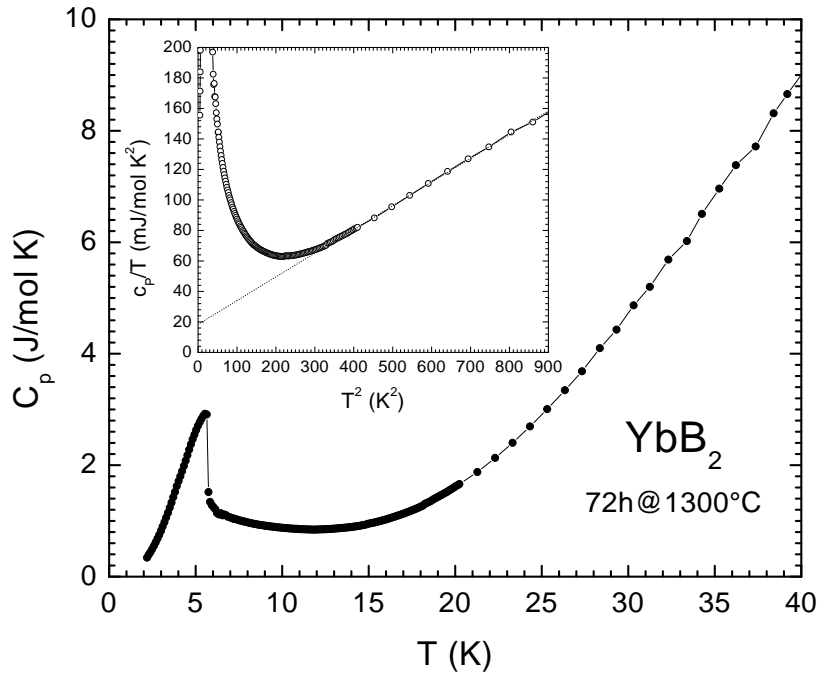}
\caption{\label{fig3} Temperature dependence of the specific heat
of a pellet from batch 1300/72, showing a sharp rise below 5.7~K
due to magnetic ordering. The inset shows the $T^2$ dependence of
$C_p/T$, where an estimate of $\gamma=18$~mJ/mol K$^2$ was
obtained by linear fitting of the data between 400 and 900~K$^2$.}
\end{figure}

Figure~\ref{fig3} shows zero-field specific heat data of sample
1300/72, taken between 2~K and 40~K. The higher temperature region
is dominated by the electronic and lattice contributions, whereas
the low temperature data shows a sharp increase in $C_p$ at 5.7~K
due to the anti-ferromagnetic ordering. From these data
$T_N=5.6\pm0.1$~K The inset to fig.~\ref{fig3} shows the same data
plotted as $C_p/T$ vs. $T^2$. A linear region is seen between 400
and 900~K$^2$, which extrapolates to $\gamma=18$~mJ/mol~K$^2$. The
obtained value should be understood only as a semi-quantitative
estimate of the actual electronic coefficient of the specific heat
in YbB$_2$ since, besides the presence of a secondary phase, the
temperature region used for the linear fit is rather high. It is
inappropriate to use the data below T$^2$ = 400~K though, where
magnetic states begin to strongly influence the measured specific
heat.

\subsection{Resistivity}

\begin{figure}[htb]
\includegraphics[angle=0,width=32pc]{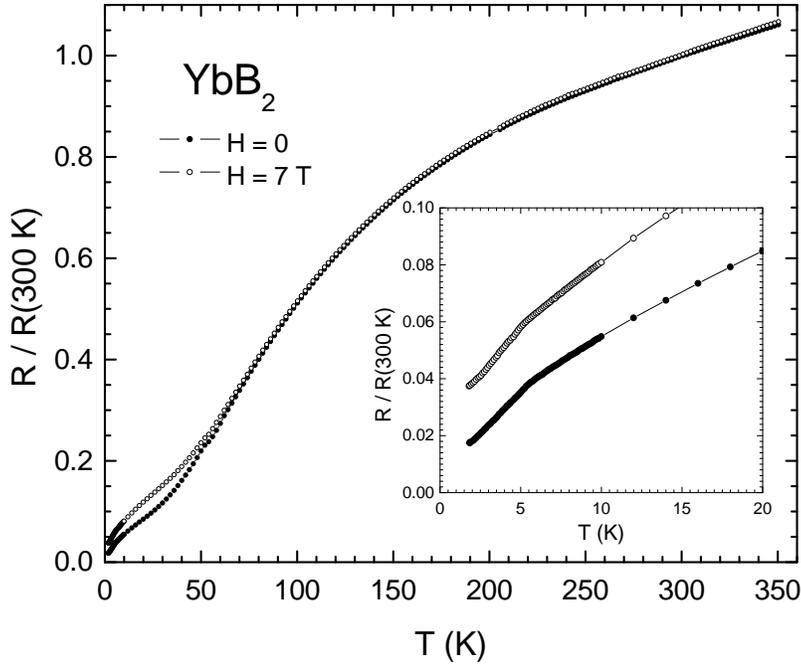}
\caption{\label{fig4} Temperature dependence of the normalized
resistance of a pellet from batch 1300/72, for $H=0$ and $H=70$
kOe. The inset details the slope change in resistance below 5.6~K
and 5.4~K respectively, due to magnetic ordering.}
\end{figure}

Figure~\ref{fig4} shows resistance data normalized to $R$(300~K),
for zero applied field and for H = 70 kOe. These measurements
demonstrate the metallic behavior of YbB$_2$, manifesting a
monotonically decreasing resistance as the sample is cooled, and
the anti-ferromagnetic ordering is accompanied by a small but
abrupt change in slope as the sample is cooled below 5.7~K,
detailed in the inset of fig.~\ref{fig4}. Analysis of $dR/dT$ for
these curves shows that $T_N(H=0)=5.6\pm0.1$~K and $T_N(H=70$~kOe$
)=5.4\pm0.1$~K, so the sppression of $T_N$ by an applied field is
small.

The broad curvature at intermediate temperatures is a common
feature of Yb compounds attributed to CEF effects, since electron
scattering is reduced as the upper CEF energy levels become
depopulated with cooling. The zero-field residual resistivity
ratio, defined here as $RRR=R(300)/R(1.8)$, is 57 for sample
1300/72 and grew along the low-pressure synthesis series, starting
at $RRR=15$ for sample 950/4. This is again consistent with the
trends seen in X-Ray diffraction and magnetization, but in the
case of transport measurements there is a much greater number of
factors that may influence the actual values, such as the
conducting properties of each phase as well as the nature/strength
of the intergrain coupling throughout the sample.

\section{Characterization of samples prepared using the high-pressure synthesis}

\subsection{X-Ray Powder Diffraction}

\begin{figure}[htb]
\includegraphics[angle=0,width=32pc]{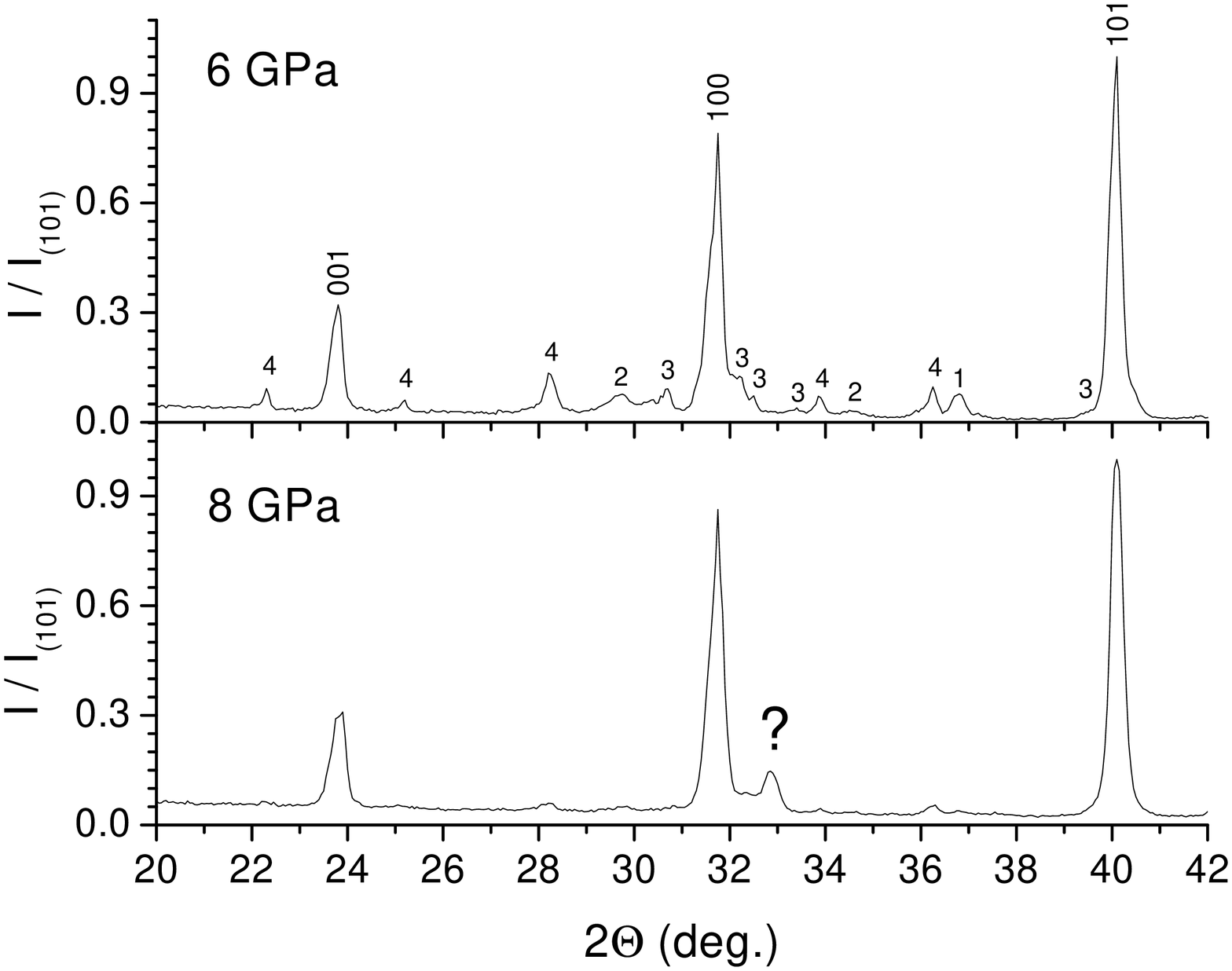}
\caption{\label{fig5} X-Ray diffraction pattern of YbB$_2$ powder
from samples 6 GPa and 8 GPa. The 3 major peaks are the YbB$_2$
reflections in the shown interval (001,100,101), and the other
labels in the top graph mark the peak positions for (1)YbO, (2)
Yb$_2$O$_3$, (3)Yb$_3$O$_4$ and (4)YbB$_4$.}
\end{figure}

X-Ray diffraction measurements for the 2 HPS samples are presented
in figure~\ref{fig5} for
$20^{\circ}\leqslant2\Theta\leqslant42^{\circ}$. The top
diffractogram, for sample 6 GPa, shows the three expected YbB$_2$
reflections in this range, but it is also possible to identify 4
impurity phases: (1)YbO, (2)Yb$_2$O$_3$, (3)Yb$_3$O$_4$ and
(4)YbB$_4$. The sesquioxide Yb$_2$O$_3$ is the standard result of
oxidization of elemental Yb, and it is interesting to note that
YbO and Yb$_3$O$_4$ are known to form by reaction of the
sesquioxide with Yb at high pressure\cite{leg78a,leg78b}. The
presence of the latter two oxide phases indicates that either the
starting elemental Yb was partially oxidized, or that there was
still some free oxygen gas present during the synthesis procedure.
Also, these X-Ray measurements were done about 6 months after the
6 GPa sample was grown and measured, and it had notably lost much
of its cohesion by this time, which may indicate that further
oxidization occurred in this period.

The lower diffractogram in fig.~\ref{fig5} is for the 8 GPa
sample, and here we observe that the previously identified
impurity phases have all but vanished, although by magnifying the
noise region it is still possible to detect the secondary phases
as being present in very small amounts, most likely not exceeding
5\% altogether. However, a new peak has appeared at
$2\Theta=32.8^{\circ}$, which was not observed in any of the
previous runs. We were unable to positively associate this peak
with any of the more likely candidate phases, such as pure
elemental ytterbium and boron, other borides and oxides, or
compounds that could possibly result from reaction of the Yb-B
melt with NaCl in the pressure cell. But, as will be discussed
below, this second phase is very likely a moment bearing one. The
refined lattice parameters for YbB$_2$ in this sample are
$a=3.239(3)$\AA, $c=3.722(4)$\AA.

\subsection{DC Magnetization}

\begin{figure}[htb]
\includegraphics[angle=0,width=32pc]{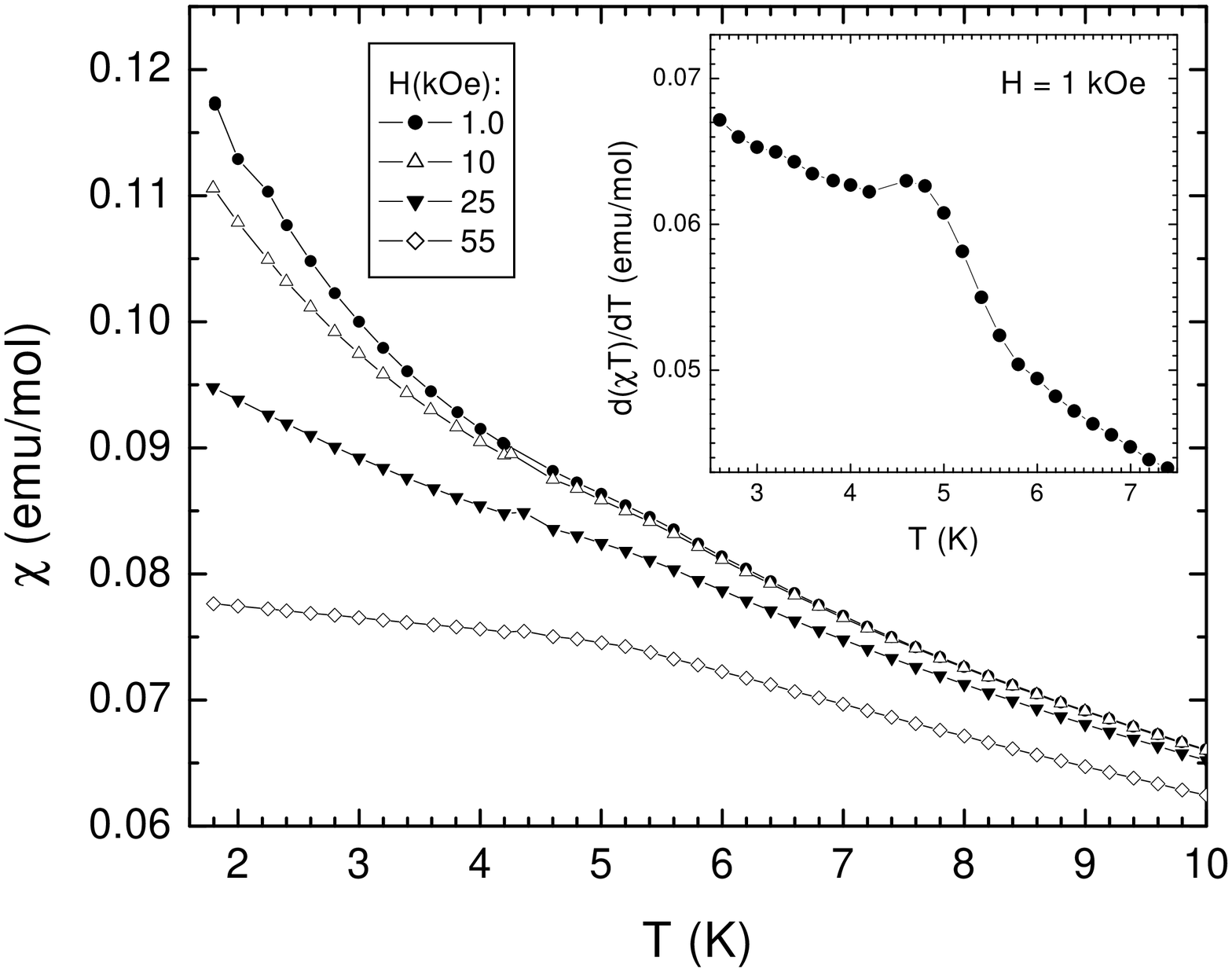}
\caption{\label{fig6} Temperature dependence of the magnetic
susceptibility of sample 8 GPa, in several applied fields. The
inset details the peak in $d(\chi T)/dT$ used to evaluate the
ordering temperature.}
\end{figure}

The magnetization curves of the two HPS samples showed influence
of secondary magnetic compounds, among which are Yb$_2$O$_3$,
known to order anti-ferromagnetically at 2.3~K\cite{moo68a,li94a},
and Yb$_3$O$_4$, reported as weakly paramagnetic\cite{leg78b} but
whose magnetic properties are as yet poorly explored. Thus, the
beginning of the magnetic ordering of YbB$_2$ at 5.7~K is not seen
as a peak, but rather as a feature occurring on top of a
paramagnetic tail. In fig.~\ref{fig6} we present the low
temperature susceptibility behavior of the 8 GPa sample for
several fields up to 55 kOe. Applying higher fields to the 8 GPa
sample tends to reduce the influence of the underlying
paramagnetic tail while showing very little effect on the ordering
temperature, and at H = 55 kOe the magnetically ordered region is
almost levelled. The inset shows the temperature derivative
$d(\chi T)/dT$ for H = 1 kOe, which gives $T_N=4.7\pm0.2$~K, a
value that is probably artificially low due to the signal from the
magnetic secondary phase contaminants.

By fitting the high temperature inverse susceptibility we found
$\mu_{eff}$ = 4.0 and 4.5~$\mu_B$/YbB$_2$ for the 6 GPa and 8 GPa
samples respectively. The deviation from 4.5~$\mu_B$/Yb for the 6
GPa sample is easily understood considering the presence of
non-magnetic phases such as YbO, YbB$_4$ and possibly unreacted
Yb. On the other hand, the 8 GPa sample displays the expected
effective moment for Yb$^{3+}$. It should be noted that the
obtained value cannot be used by itself as a demonstration of the
purity of YbB$_2$ since, coincidentally, Yb$_2$O$_3$ has almost
exactly the same molecular mass per Yb atom as YbB$_2$ (197.04 and
195.04 g/mol(Yb) respectively), but when combined with the X-Ray
results it becomes consistent with a much improved phase purity.

\subsection{Heat Capacity}

\begin{figure}[htb]
\includegraphics[angle=0,width=32pc]{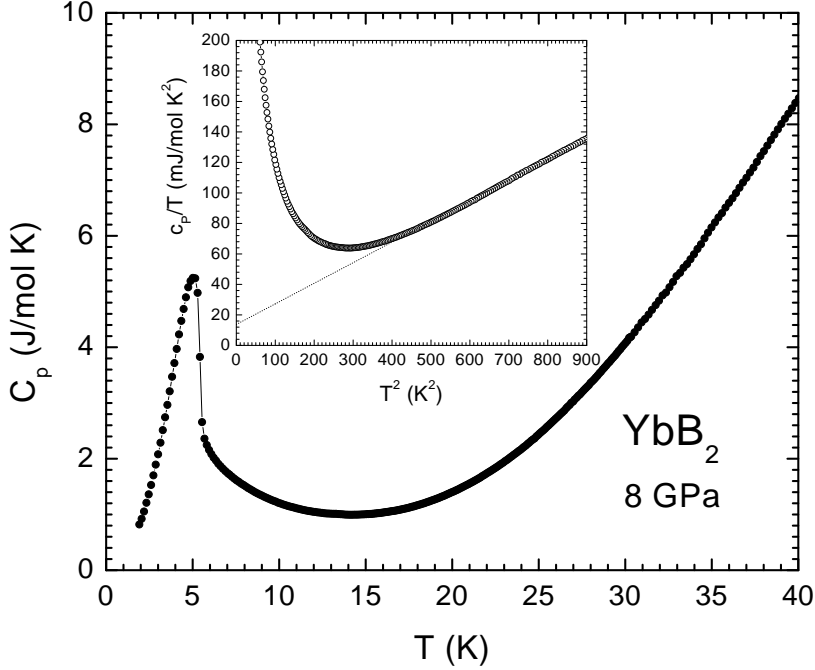}
\caption{\label{fig7} Temperature dependence of the specific heat
of sample 8 GPa. The inset shows the $T^2$ dependence of $c_p/T$,
where an estimate of $\gamma=14$~mJ/mol~K$^2$ was obtained by
linear fitting of the data between 400 and 900~K$^2$.}
\end{figure}

Figure~\ref{fig7} shows zero-field specific heat data of the 8 GPa
sample, taken between 2~K and 40~K. The higher temperature
behavior (dominated by the electronic and lattice contributions)
is very similar to that observed in fig.~\ref{fig4} for the LPS
sample, but as we cool below $\sim15$~K the $C_p$ values of the
HPS sample rises much faster, reaching a value of 5.2 J/mol~K at
the peak, as compared to the 3.0 J/mol~K for sample 1300/72. This
is qualitatively consistent with the fact that the 8 GPa sample
should have a greater mass proportion of YbB$_2$. The feature in
$C_p$ at the ordering temperature is broader in the 8 GPa sample
though, so our best estimate is $T_N=5.4\pm0.2$~K from these data.

The inset shows the same data plotted as $C_p/T$ vs. $T^2$, where
the extrapolation of the linear region is between 400 and 900
K$^2$ results in $\gamma=14$~mJ/mol~K$^2$, close to the value
obtained for sample 1300/72. It is worth noting that these values
are significantly larger than those obtained both theoretically
and experimentally for many other members of the metal-diboride
family\cite{vaj01a}, and indicates that at least some
hybridization of the ytterbium $4f$ levels with the conduction
band electrons may be occurring, with consequent heavy-electron
behavior. The study of pressure effects in the specific heat of
this compound may provide further evidence of such behavior.

\subsection{Resistivity}

\begin{figure}[htb]
\includegraphics[angle=0,width=32pc]{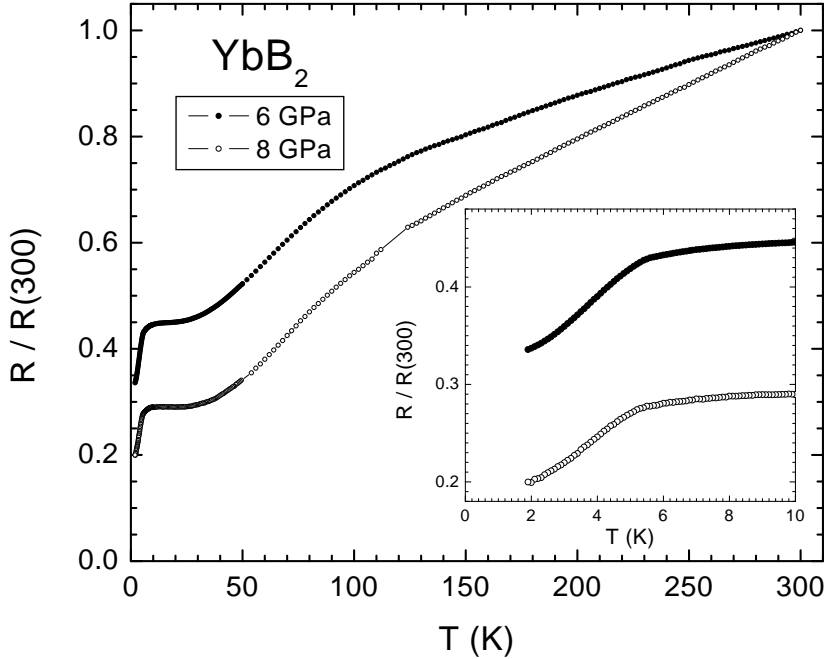}
\caption{\label{fig8} Temperature dependence of the normalized
resistance for samples 6 GPa and 8 GPa. The inset details the drop
in resistance below 5.7~K due to magnetic ordering.}
\end{figure}

Figure~\ref{fig8} shows zero-field resistance data normalized to
$R$(300~K), for the 6 GPa and 8 GPa samples. The general features
observed in the LPS samples are also present here, such as
monotonic decrease on cooling, broad curvature at intermediate
temperatures, and drop in resistivity at the ordering temperature.
The main difference is the almost levelled resistance behavior
below $\sim30$~K, persisting until the magnetic ordering occurs,
at which point resistance begins to drop quite sharply. From a
clearly marked discontinuity in $dR/dT$ we obtain
$T_N=5.7\pm0.1$~K for both samples, although the slope continues
to rise until much lower temperatures. The $RRR$ values obtained
for the 6 GPa and 8 GPa samples are only 3 and 5 respectively.

\section{Discussion}
\label{sec:discussion}

We can now compare the results obtained for the LPS and HPS
samples. The X-Ray diffractograms allowed us to identify most of
the secondary phases present in each sample resultant from these
reaction routes. The sample that resulted closest to single-phase
YbB$_2$ was the one where the elements were melted and
recrystallized at 8 GPa. However, this sample contained several
different impurity compounds in small quantities, including one
that remained unidentified. In contrast, the best sample obtained
by the LPS route (1300$^{\circ}$C for 72 hours) still contained a
significant amount of YbB$_4$ as a secondary phase which, however,
was the only other compound detectable in the diffractogram.

Magnetization measurements showed that all samples ordered
anti-ferromagnetically below 5.7~K regardless of preparation
method, with the estimated values from $d(\chi T)/dT$ being
$T_N=5.4\pm0.2$~K for the 1300/72 sample, and $T_N=4.9\pm0.2$~K
for the 8 GPa sample. This latter value is in disagreement from
those obtained by all other experiments, and probably a
consequence of interference in the $d(\chi T)/dT$ analysis by the
secondary magnetic phases. The calculated effective moment per Yb
atom was closest to its expected value of 4.5~$\mu_B$ for the 8
GPa sample, demonstrating that the Yb atoms are in or very close
to their 3+ state, and the higher phase purity in this sample.
However, the fact that the 1300/72 sample contained only a
non-magnetic compound as impurity allowed a much clearer
observation of the drop in susceptibility resultant from
anti-ferromagnetic ordering. Deviations from Curie-Weiss behavior
below $\sim$100~K were observed in all samples and likely caused
by CEF effects on the Yb electronic energy levels.

Specific heat measurements allowed a better precision in the
estimation of the ordering temperature for both the 8 GPa and the
1300/72 samples, resulting in $T_N=5.4\pm0.2$~K and
$T_N=5.6\pm0.1$~K respectively. These samples showed an electronic
specific heat coefficient in the range of 14 and 18~mJ/mol~K$^2$
respectively, indicating a possibly enhanced electron mass
character of the conduction electrons. Our lack of single-phase
samples and of an equivalent non-magnetic compound (such as
LuB$_2$) to discount the electron/lattice contributions prevents
us from attempting any reliable analysis of the magnetic entropy
behavior, but it is clear that the 8 GPa sample has accumulated
much more magnetic entropy up to the transition temperature.

Resistance measurements showed monotonically decreasing metallic
behavior in all samples when cooled from room temperature,
including a broad curvature at intermediate temperatures often
seen in Yb compounds with CEF splitting. The anti-ferromagnetic
ordering was marked as a sudden change in the zero-field
resistance slope due to the decrease of spin disorder, at
$T_N=5.7\pm0.1$~K for the HPS samples and $T_N=5.6\pm0.1$~K for
the LPS sample. A small but noticeable decrease to
$T_N=5.4\pm0.1$~K was seen in the resistance measurement at
$H=70$~kOe for the latter sample. The HPS samples showed much
higher electron scattering and lower residual resistivity ratios,
possibly due to a greater amount of lattice defects and/or to the
influence of non-conducting impurity phases most likely present as
grain boundaries. However, one could argue that the better
conducting properties of the LPS samples were influenced by the
presence in larger proportion of YbB$_4$ which is known to be
metallic\cite{eto79a}, and therefore the actual transport behavior
of YbB$_2$ remains to be determined more accurately, if
single-phase samples and possibly single crystals are developed in
the future.

\section{Conclusion}
\label{sec:conclusion}

In this paper we presented a characterization of some
thermodynamic and transport properties of sintered YbB$_2$ pellets
prepared by low-pressure and high-pressure synthesis methods. The
two different routes showed several similarities that could be
attributed to this compound, but also several relevant differences
in sample behavior that arise from the nature and amount of
secondary phases. Based on these data we can infer that in YbB$_2$
the Yb atoms are essentially in their 3+ state and order
magnetically below $T_N=5.6\pm0.2$~K, which is a high ordering
temperature for an Yb-based intermetallic compound. The magnetic
ground state is very likely anti-ferromagnetic. We have also
obtained indications of CEF effects and enhanced electron mass at
the Fermi level. Further development towards obtaining pure,
single-phase samples (and if possible in single crystalline form)
would be very helpful in corroborating or correcting the results
presented here, but the fact that two such different techniques
gave robust results gives us some confidence in our basic
understanding of the magnetic and electronic properties of
YbB$_2$.

Ames Laboratory is operated for the US Department of Energy by
Iowa State University under Contract No. W-7405-Eng-82. This work
was supported by the Director for Energy Research, Office of Basic
Energy Sciences.










\end{document}